# Enhancing the robustness of scale-free networks

**Jichang Zhao and Ke Xu**
State Key Laboratory of Software Development Environment
Beihang University, 100191, Beijing, P.R.China
Email: zhaojichang@nlsde.buaa.edu.cn and kexu@nlsde.buaa.edu.cn

**Abstract**

Error tolerance and attack vulnerability are two common and important properties of complex networks, which are usually used to evaluate the robustness of a network. Recently, much work has been devoted to determining the network design with optimal robustness. However, little attention has been paid to the problem of how to improve the robustness of existing networks. In this paper, we present a new parameter $\alpha$, called enforcing parameter, to guide the process of enhancing the robustness of scale-free networks by gradually adding new links. Intuitively, $\alpha < 0$ means the nodes with lower degrees are selected preferentially while the nodes with higher degrees will be more probably selected when $\alpha > 0$. It is shown both theoretically and experimentally that when $\alpha < 0$ the attack survivability of the network can be enforced apparently. Then we propose new strategies to enhance the network robustness. Through extensive experiments and comparisons, we conclude that establishing new links between nodes with low degrees can drastically enforce the attack survivability of scale-free networks while having little impact on the error tolerance.

## 1. Introduction

We live in a world of complex networks, e.g. the Internet, WWW, power grids, social networks and etc. Through probing, collecting and analyzing the topology data of complex networks, people found that the traditional ER (Erdos-Renyi) model [1] cannot well simulate some practical networks such as Internet and WWW. For WWW, Barabάsi and Albert found that the degree distribution follows the power-law form $p(k)=k^{-r}$ with $r > 2$. This feature also appears to exist in many other complex networks and such networks are called scale-free networks. It is interesting but somewhat amazing that Barabάsi and Albert proposed a simple mechanism, called BA model [2], to explain the generation of scale-free networks. Since then, the study of scale-free networks has attracted wide attention from many different research fields. Our group developed Dolphin System [3, 4] to probe the IPv6 Internet and found that the IPv6 AS (Autonomous System) backbone network was also scale-free with $r < 2$ [5]. The Cooperative Association for Internet Data Analysis (CAIDA) [6] also developed their own IPv6 probing tool called Scamper [7].

An important characteristic of scale-free networks is the heterogeneity of the degree distribution, which makes the scale-free network tolerant to random failures but extremely vulnerable to malicious attacks. Error tolerance and attack vulnerability are two common and important properties of complex networks [8]. The relative size of the largest cluster $S$ [8, 9] and the average inversed geodesic $L^{-1}$ [9] are used to characterize the behavior of the network during attacks. We can also introduce the average network efficiency [10]

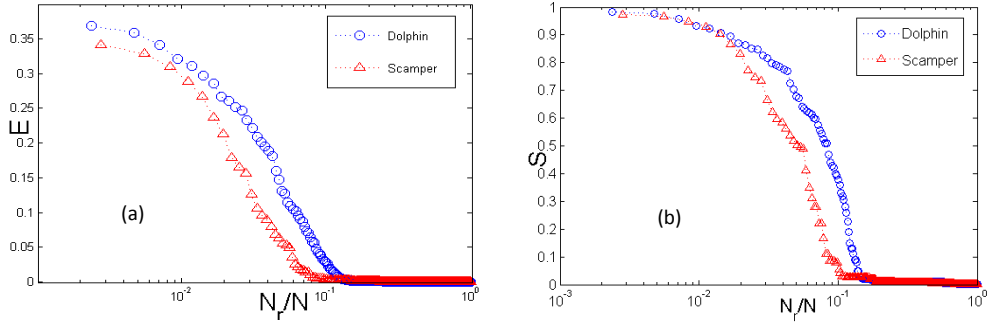

**Figure 1.** The numerical experiment is implemented on the topology data from Dolphin and CAIDA Scamper. Both (a) and (b) are linear-log plots, the horizontal axis is defined as the fraction of removed nodes, which is $N_r/N$. We use $E$, the average network efficiency in (a) and $S$, the relative size of the largest cluster in (b) to characterize the behavior during the attack. And the attack is based on the degree of the node. The results show that $f_t^{Dolphin} \approx 0.14$, $f_t^{Scamper} \approx 0.12$.

$$E = \frac{1}{N(N-1)} \sum_{i=1, j=1}^{N} \frac{1}{d_{ij}}, \quad (1)$$

where $N$ is the size of the network and $d_{ij}$ is the length of the shortest path between $i$ and $j$. The error tolerance of the network can be measured by the entropy, which is computed based on the degree distribution of the network [11, 12]. During the malicious attacks or random failures, $f_c$ is always used to characterize the critical fraction of the least nodes that need to be removed until the network is collapsed [13, 14]. $f_c^{rand}$ and $f_c^{targ}$ were used as responses to random failures and malicious attacks respectively in [13]. In particular, $f_c^{rand}$ can be easily computed by the following formula [15]

$$f_c^{rand} = 1 - \frac{1}{\frac{<k^2>}{<k>} - 1}, \quad (2)$$

where $<k>$ is the first moment (mean value) of the degree and $<k^2>$ is the second moment of the degree. But for $f_c^{targ}$, we need to solve a few difficult functions depending on the degree distribution [13, 16, 17]. Some optimal models were presented to generate robust networks against random failures and attacks [13, 14]. And it was also proved that there were no more than three node connectivities in optimal networks [18].

In fact, it is impractical to keep $<k>$ a constant and rewire the links [13] for real networks. For example, either the Internet or the power grids have been formed so long and thus it is almost impossible to re-establish them to enhance the network robustness. However, one thing we could do is to add a smaller number of links to the network to achieve higher robustness.

In this paper, we study the problem of how to improve the robustness of existing networks and find that the attack survivability of scale-free networks can be enforced greatly by gradually adding new links between the nodes with low degrees, while having little impact on the error tolerance. The rest of the paper is organized as follows. In section 2, malicious attacks based on degree are performed and the robustness of scale-free networks is analyzed. In section 3, we mainly discuss how to enforce the network robustness based on extensive numerical experiments. Theoretical analysis is given in section 4. Finally, the conclusions are summarized in section 5.

**2. Malicious attacks based on heterogeneity**

The malicious attack towards a network is based on the heterogeneity of the degree distribution. To the opposite of the random failure, it removes the most important node from the network first. We suppose the attackers know the global topology of the network, thus they can locate the key node and attack it. And in fact, this might happen in real networks. We relate the importance of the node with its degree, and in such a case, the most import one is also the one with the highest degree. Then we perform experiments of malicious attacks based on the IPv6 AS backbone network topology.

In this paper, we have two data sources from the Dolphin System and the Scamper System respectively. Both of the systems were developed to discover the global IPv6 backbone network by traceroute. The dataset from the Dolphin System has 419 nodes and 1820 edges, and the other from the Scamper System has 356 nodes and 1007 edges.

In this paper, we use $f_r$ and $f_t$ to denote responses to random failures and malicious attacks respectively. In figure 1, IPv6 AS backbone network is vulnerable under malicious attacks with $f_t \approx 0.14$, but robust to random failures with $f_r \approx 0.96$. In order to extend the analysis, we relate the importance of the node with its betweenness [19, 20, 21], and find that the result is quite close to the previous result. This is mainly due to the betweenness of the node is strongly related with its degree in the IPv6 AS backbone network without fractal character [22, 23, 24].

In practical applications, we hope that the network can be tolerant to random communicating errors or failures, and also can be robust to malicious attacks, especially in the martial field. However, the above experiments show that the attacker just need to remove a very small part of the key nodes to make the whole network collapsed and malfunctioned. In this paper, we define the robustness as error tolerance and attack survivability. In addition, high robustness means both high error tolerance and attack survivability. So we hope to enforce the robustness of scale-free networks against attacks but still keep the error tolerance.

**3. Enhancing network robustness by adding new links**

The scale-free network can be represented as an undirected graph $G(V, E)$, where $V$ is the set of nodes and $E$ is the set of links. Define $\Psi$ as the set of all the possible links between the nodes in $V$. Define the set of links in the complementary graph as

$$\overline{E} = \Psi - E.  \tag{3}$$

Define the link degree as

$$k_e = k_s k_d,  \tag{4}$$

where $k_s$ and $k_d$ are the degrees of two nodes of the link $e$. We define the link degree as the product of the degrees of the nodes it connects mainly because the product is strongly related to the betweenness of the link [9], which can be used to characterize the importance of the link. We propose a new parameter $\alpha$, called enforcing parameter, and define the probability of choosing a new link $e_i$ from $\overline{E}$ as

$$p(k_{e_i}) = \frac{k_{e_i}^{\alpha}}{\sum_{e_j \in \overline{E}} k_{e_j}^{\alpha}}. \tag{5}$$

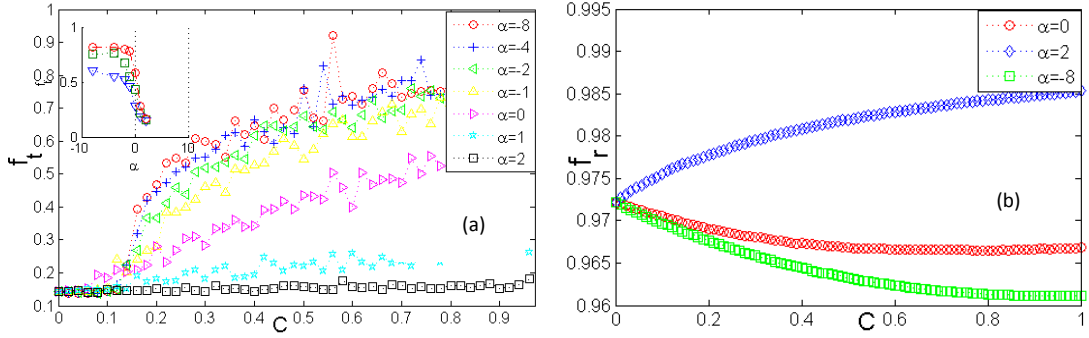

**Figure 2.** We enforce the network robustness with the topology data from Dolphin under different $\alpha$. The horizontal axis is defined as the cost $C$. In (a), $\alpha$= -8,-4,-2,-1,0,1,2 from top to bottom, and it is easy to find that the effect of the enforcement with the decrease of $\alpha$ becomes more and more apparent, while $f_t$ keeps unchanged when $\alpha = 2$. In the inner chart, it shows different $f_t$ after enforcement under different $\alpha$ with $C$=0.3, 0. 5, 1 from top to bottom. It can be found that the network robustness can be enforced more strongly when $\alpha$ decreases under the same cost. In (b), it shows the fluctuation of $f_r$ during the enforcement. It is easy to find that $f_r$ increases when $\alpha = 2$ and decreases when $\alpha$=0 or -8, but in a little range, e.g. $\Delta f_r \approx 0.01105$ when $C$=1.

The cost is a key factor that constrains the structure in establishing a network. We define the cost $C$ as the ratio of the number of new links to the number of links in the initial network, that is

$$C = \frac{|E_{new}|}{|E_{init}|}. \tag{6}$$

Define

$$\kappa = \frac{<k^2>}{<k>}. \tag{7}$$

From (2), we can compute $f_r$ using the formula

$$f_r = 1 - \frac{1}{\kappa - 1}. \tag{8}$$

But for $f_t$, we need to check whether the condition $\kappa < 2$ is satisfied. When it is satisfied, the critical fraction will be $f_t$ [13]. By use of $f_r$ and $f_t$, we can characterize the error tolerance and the attack survivability of the network directly and numerically.

First, we perform numerical experiments of enhancing robustness on IPv6 AS backbone network topologies obtained respectively from the Dolphin System and the Scamper System. We perform the experiments with each pair of $\alpha$ and cost for 100 times and get the mean value as the final result. We also assume the data obeys the Gaussian distribution and then get the two-sided confidence interval with confidence coefficient equal to 95%. For example, the two-sided confidence interval for $f_t$ is [0.710627, 0.727702] and finally we get $f_t = 0.719165$ when $\alpha = -8$, $C=0.5$. We compare the experimental results under different values of $\alpha$. In figure 2, we find when $\alpha < 0$, the attack survivability can be greatly enforced and becomes even better as $\alpha$ decreases. When $\alpha = 0$, the attack survivability is improved slowly and the speed becomes even slower as $C$ increases. There is no apparent tendency of improvement when $\alpha > 0$. In this case, the attack survivability just fluctuates a little above the initial state and begins to keep almost unchanged as $\alpha$ increases.

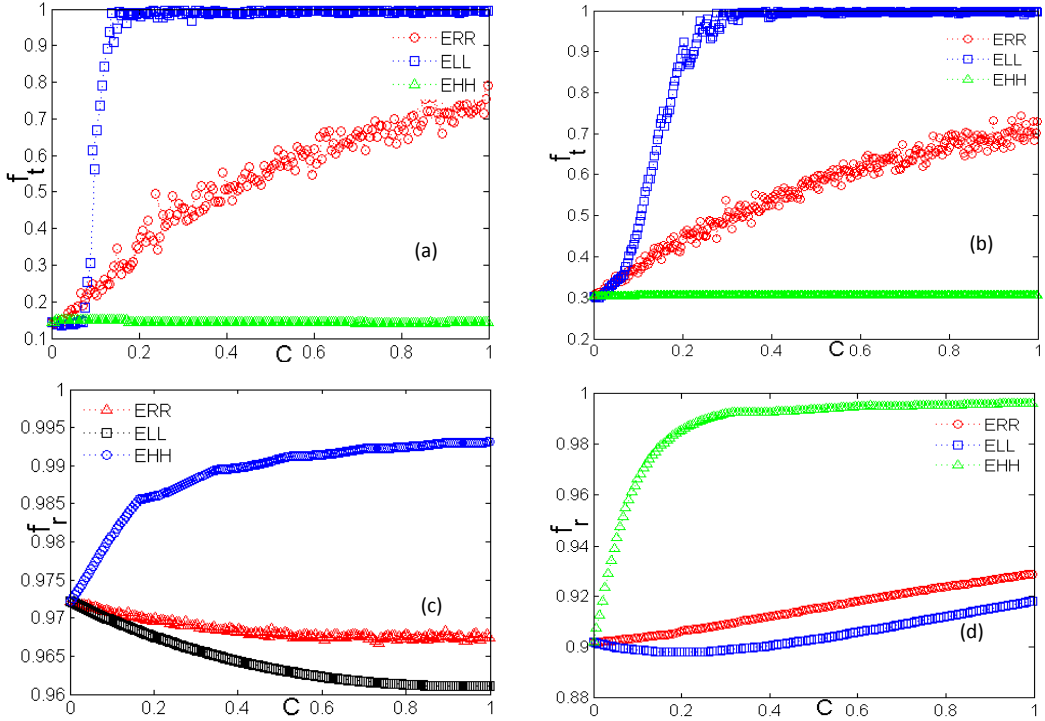

**Figure 3.** We enforce the network robustness with the topology data from Dolphin and *BA* (3, 1000) under the strategy of ERR, ELL and EHH respectively, using $f_t$ and $f_r$ to characterize the fraction. (a), (c) is the results based on the data from Dolphin while (b), (d) based on the data from *BA* (3, 1000). It is shown from (a) and (b) that ELL strategy can enforce the network robustness drastically: $C_t^{Dolphin} \approx 0.17$ and $C_t^{BA} \approx 0.29$ when $f_t \approx 1$. From (c) and (d), it also demonstrates that $f_r$ just fluctuates in a little range: $\Delta f_r^{Dolphin} \approx 0.004$ when $C = C_t^{Dolphin}$ and $\Delta f_r^{BA} \approx 0.003$ when $C = C_t^{BA}$.

In order to analyze the evolving of error tolerance when adding new links, we compute $f_r$ as a function of the cost $C$. In figure 2 (b), it can be found that $f_r$ decreases a little when $\alpha \leq 0$ (in the range of 0.01), while increases when $\alpha > 0$. We can conclude from the above numerical experiments that the survivability of the network can be enforced without apparent impact on the error tolerance when $\alpha < 0$, and especially when $\alpha$ decreases, while the situation is quite different when $\alpha > 0$. In fact, $\alpha = 0$ means selecting two nodes randomly from the network which are not connected to each other and establishing a new link between them; $\alpha < 0$ means the nodes with lower degrees are selected preferentially while the nodes with higher degrees will be more probably selected when $\alpha > 0$. In order to find a more efficient way to enforce the attack survivability of the network, we let $\alpha = 0$, $\alpha = -\infty$ and $\alpha = +\infty$ respectively, then we can get three different enforcing strategies:

- ERR: Select a pair of isolated nodes randomly in the network and establish a new link between them.
- ELL: Select a pair of isolated nodes with the lowest degree in the network and establish a new link between them.
- EHH: Select a pair of isolated nodes with the highest degree in the network and establish a new link between them.

In figure 3 (a), it is easy to find that ELL can enforce the attack survivability of the network with a low cost while EHH has little effect and ERR can only improve the attack survivability slowly and become even slower as $C$ increases. At the same time, though the error tolerance of the network is weakened by ELL and ERR, but the impact is slight because $f_r$ is still more than 0.9. We can conclude that the attack survivability and the error tolerance are mutually exclusive to each other based on the above analysis of experiments. Because of the nature of heterogeneity, there are a lot of nodes with low degree in the network, which are called edge nodes, but a few nodes may have very high degrees, which are called hub nodes. So the network is robust to random failures and vulnerable to malicious attacks. Therefore, in order to enhance the network robustness, we need to make a trade-off between the attack survivability and the error tolerance. From the above analysis, we know that the ELL strategy is a good way to enhance the network robustness because it can drastically improve the attack survivability while keeping the high error tolerance almost unchanged as the cost $C$ increases.

We also perform experiments on BA model to verify the above conclusion. The BA model is a model of a growing network which starts from $m_0$ nodes and adds new nodes preferentially connecting to the existing nodes. In this paper, we use $BA(m, N)$ to represent the BA model, where $N$ is the size of the generated network and a new node is added to the network with preferential links to $m$ existing nodes. In figure 3 (b), we can find that the ELL strategy is still quite effective on the BA model. It means that we can extend our conclusion from the scale-free network with $r < 2$ to the ones with $2 < r \leq 3$.

In fact, the clustering coefficient of one node in a network represents the closeness of its neighbors. In scale-free networks, the clustering coefficient of a hub node is low determined by the disassortative property. Once a hub node is attacked, its neighbors with low degrees would collapse for losing central transitive node. We can establish new links among the edge nodes which are the neighbors of the hub node to form a local loop. Because of the local loop, edge nodes can still connect to each other even when the hub node is attacked and collapsed. In the ELL strategy, the new links added to the network are mainly between the nodes with low degrees, so their degrees increase. However, they still occupy a large part. In contrast, the nodes with high degrees keep unchanged with few new links connecting to them. This can be found in figure 4 (a).

**4. Theoretical analysis**

To verify our conclusion above, we also provide theoretical explanation here. In order to simplify the analysis, we assume that the new links added to the network are assigned at first. We define the fraction of new links for each node in the network is $r_i$, where $N$ is the size of the network and $1 \leq i \leq N$. Then we can compute $r_i$ using the following equation

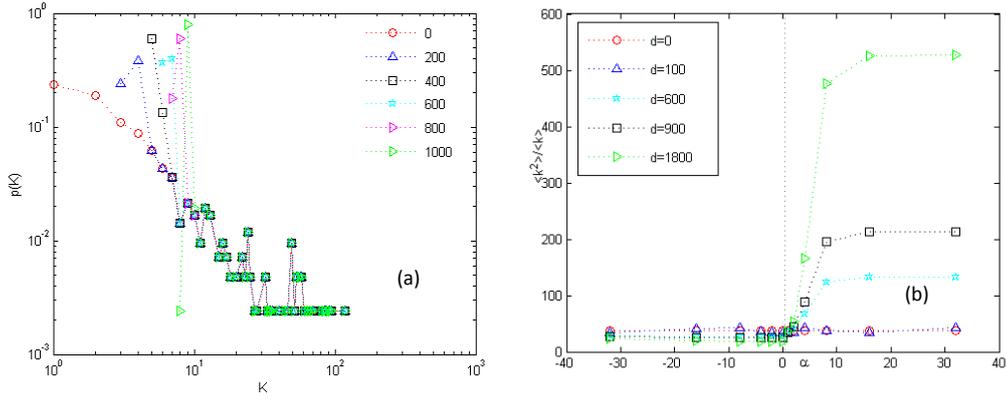

**Figure 4.** (a) is the degree distribution of the topology data from Dolphin during the enforcement of robustness with the ELL strategy, where $d$ =0, 200, 400, 800 and 1000 from top to bottom. (b) is the variation of $\kappa$ under different $\alpha$, where $d$ =0,100,600,900 and 1800 from top to bottom.

$$r_i = \sum_{j=1}^{N} \frac{k_i^\alpha k_j^\alpha}{\sum_{p,q=1} k_p^\alpha k_q^\alpha}, \qquad (9)$$

where $i \neq j$, $p \neq q$, $i$ is not connected to $j$, and $p$ is not connected to $q$. Assume that the cost is $C$, therefore we have $d = C/|E|$ new links added to the network. We can also use the condition $\kappa < 2$ to check whether the network is collapsed. From the above analysis, we can easily obtain from (7) that

$$\kappa = \frac{<k^2>}{<k>} = \frac{\frac{1}{N}\sum_{i=1}^{N}(r_i d + k_i^0)^2}{\frac{1}{N}2(d+|E|)}, \qquad (10)$$

where $k_i^0$ is the degree of $i$ in the initial network. Figure 4 (b) shows the variation of $\kappa$ based on equation (10). It is easy to find that $\kappa$ increases when $\alpha > 0$ and keeps almost unchanged when $\alpha < 0$. The reason is that when $\alpha > 0$, $r_i$ grows when $k_i^0$ increases. Thus $\kappa$ will increase when adding $d$ new links. In contrast, when $\alpha < 0$, $r_i$ grows when $k_i^0$ decreases, the variation of $\kappa$ will be smoothed. Then according to (8), $f_r$ will increase when $\alpha > 0$ but keep almost unchanged when $\alpha < 0$, which is similar to the result of the above numerical experiments. As for $f_t$, it will increase because the heterogeneity of the network is weakened by adding new links when $\alpha < 0$.

## 5. Conclusions

It has been proved that the scale-free network is robust to random failures but vulnerable to malicious attacks. However, in practical applications we hope that the network can be robust against not only inevitable errors during the communication but also malicious attacks, especially in the martial field. One possible way to enhance the robustness of an existing network is to add new links to it. In this paper, we propose a new parameter $\alpha$, called enforcing parameter, to guide the process of enhancing the robustness of scale-free networks by gradually adding new links and the experiments show that the effect of enhancement is better when $\alpha < 0$. Then, three enforcing strategies: ERR, ELL and EHH are presented. Through experiments and analysis, it is shown that the ELL strategy can greatly enforce the attack survivability of the network without negative effect on the error tolerance. In summary, we can efficiently enhance the robustness of scale-free networks by adding new links under the ELL strategy.

## Acknowledgements

We would like to thank two anonymous reviewers for their valuable comments and suggestions which helped to improve the paper significantly. This work is supported by National 973 Program of China (Grant NO. 2005CB321901). We thank Maksom Kistak for the suggestion on how to compute the betweenness and Tomas Manke for the discussion of entropy in complex networks.